\documentclass[12pt]{iopart}
\usepackage{amsmath}
\usepackage{amsfonts}
\usepackage{amssymb}
\usepackage{graphicx}
\usepackage{color}

\usepackage{amssymb,amsthm}
\usepackage{amsmath}
\usepackage{subcaption}
\usepackage{graphicx}
\usepackage{color}
\usepackage{ulem}
\usepackage{url}
\usepackage{lineno}
\usepackage{arydshln}
\usepackage{multirow}
\usepackage{here}
\usepackage{bbm}

\usepackage{tikz}
\usetikzlibrary{arrows.meta, positioning, shapes.geometric, fit, backgrounds}
\usepackage{hyperref}

\begin{document}

\title[]{Characterization of the Scattered Light Noise in KAGRA Interferometer}

\author{Shih-Hong Hsu$^{1}\footnote{corresponding author}$, Hirotaka Yuzurihara$^{2,3}$, Chia-Jui Chou$^4$, Yi Yang$^4$}

\address{$^1$ Department of Electrophysics, National Yang-Ming Chiao-Tung University, Hsinchu 300, Taiwan}
\address{$^2$ Institute for Cosmic Ray Research (ICRR), KAGRA Observatory, The University of Tokyo, Kamioka-cho, Hida City, Gifu 506-1205, Japan}
\address{$^3$ National Institute of Technology, Oshima College, Komatsu, Suo-oshima, Yamaguchi, 742-2193, Japan}
\address{$^4$ School of Physical Science and Technology, ShanghaiTech University, China}

\ead{shihhonghsu@gmail.com}
\ead{yuzu@icrr.u-tokyo.ac.jp}
\ead{zhoujr@shanghaitech.edu.cn}
\ead{yangyi3@shanghaitech.edu.cn}

\begin{abstract}
In a gravitational wave (GW) telescope, non-stationary noise can prevent stable interferometer operation and limit the GW analyses. 
Among the non-stationary noise, we focus on the scattered light noise. It fluctuates across a wide frequency band within a short time. 
Therefore, the scattered light noise can mimic the GW signal, increase the false alarm rate, or bias the parameter estimation results.
In this work, we propose a new analysis method, named f-cluster, to characterize these features without prior assumptions about the scattering source.
The scattered light noise shows characteristic arch-shaped features in time-frequency maps. Therefore, the algorithm extracts the periodicity of the arch shapes from the spectrogram.
The algorithm detects the occurrence times and periodicity of excess-power regions in the spectrogram to estimate the oscillation frequency of the scattering object.
To demonstrate the algorithm, we applied this method to KAGRA data from April 1 to 30, 2025, and successfully identified a scattering frequency of $f_{sc} = 0.432$ Hz. 
Furthermore, we found a strong correlation between the occurrence rate of the scattered light noise and the amplitude of ground motion in the $0.3-1.0$ Hz band with a Pearson correlation coefficient of 0.82.
This correlation showed that the occurrence of scattered light noise can be forecast from the seismic motion.
We confirmed that our algorithm identifies the frequency of the scattered light noise and its occurrence time, without making assumptions about the scattering object, thereby contributing to the characterization of the scattered light noise in real data.
Our algorithm provides information that leads to further investigation into the noise hunting to specify the source of scattered light noise.
\end{abstract}

% \subjectindex{xxxx, xxx}

\maketitle
% \tableofcontents

%%%%%%%%%%%%%%%%%%%%%%%%
\section{Introduction}
% 1. Review on O1 ~ O4 and important events, GW150914, GW170817, etc.
Gravitational waves (GWs) are the ripples in spacetime predicted by Einstein's general relativity \cite{Einstein:1918}.
The first GW event from the merger of two black holes (named GW150914) was directly detected on September 14, 2015, marking the beginning of GW astronomy \cite{Abbott:2016}.
On August 17, 2017, the first GW event from the merger of two neutron stars (named GW170817) was detected \cite{Abott:2017}.
With the electromagnetic wave counterparts \cite{Abott_g:2017}, this observation marked the beginning of the multi-messenger era with GWs \cite{Abott_m:2017}.
A global GW network consists of two LIGO \cite{Gregory:2010, aLIGO:2020} in the United States; Virgo \cite{Acernese:2014} in Italy; GEO 600 \cite{Willke:2002} in Germany; and KAGRA \cite{Akutsu:2018} in Japan. 
With ongoing telescope upgrades and the planned addition of the Einstein Telescope \cite{Punturo:2010} and LIGO-India \cite{LIGO_I:2013}, the global GW network is moving toward higher sensitivity and broader sky coverage, establishing GW astronomy as a key tool for multi-messenger astrophysics.
The global GW network of observatories has conducted four major observing runs.
The fourth observing run, O4, is the first joint observation involving LIGO, Virgo, and KAGRA \cite{LVK:2025}.

% 3. Introduction to Glitches, GravitySpy, how glitches affect low-latency detection
Strain data from the GW telescope contain various types of noise, which can be characterized as either stationary or non-stationary.
Narrow-band stationary noise is referred to as line noise, which limits long-duration GW searches, while broadband stationary noise degrades the sensitivity and the performance of all types of GW searches.
Non-stationary noise typically lasts for a few seconds and spans a broad frequency range.
Such non-stationary noise can prevent stable interferometer operation and also mimic GW signals. 
When analyzing data containing non-stationary noise, gravitational wave searches can produce false positives, thereby reducing the performance of the transient GW search. 
At the same time, bias in the parameter estimation of GW candidate events is reported \cite{Powell:2018}.
For example, during the GW170817 event, non-stationary noise occurred at the LIGO Livingston.
To characterize and extract the non-stationary noise, the LIGO-Livingston data required additional treatment before analysis. \cite{Pankow:2018}.
In the future observing run, the extra time required by these treatments could delay sky localization.

%%% general introduction of Q-transform
To investigate and characterize noise, spectrograms are commonly used to visualize the noise features in both the time and frequency domains.
The constant Q-transform is a modification of the short-time Fourier transform.
Based on the different quality factor Q, the Q-transform provides varying resolutions in time and frequency.

There is a variety of non-stationary noise in the GW telescope. 
The Gravity spy is a project aimed at classifying non-stationary noise using the Q-transform and machine learning \cite{Zevin:2017}. 
The classification algorithm in the Gravity Spy project is trained on human-classified examples of noise in the Q-transformed time-frequency maps.

% 4. Brief introduction on the scattered light noise of KAGRA. "This is the first paper studying KAGRA's scattered light noise."
Scattered light noise is one type of non-stationary noise observed in GW telescopes.
The mechanism that induces scattered light noise is well modeled. 
In the cavities of the interferometer, when light is back-scattered into the main optical path due to reflections on scattering objects, the motion of the scattering object up-converts to the scattered light noise \cite{Accadia:2010zzb}.
A feature of the scattered light noise is the characteristic periodic, arch-shaped excess power regions in time-frequency maps.
Figure \ref{fig:SL_example} shows the characteristic arch shapes of the scattered light noise in a time-frequency map.
Despite the distinctive pattern in the time-frequency maps, manual identification is inefficient due to its up-conversion nature.
The scattered light noise has been reported as an issue in gravitational-wave telescopes such as Virgo \cite{Accadia:2010zzb, Longo:2022, Longo:2023} and LIGO \cite{Guillermo:2017, David:2021, Soni:2021, Soni:2025}, and various analysis methods have been proposed.

%%% intro. of the old method
The source of scattered light noise reported in GW telescopes varies by telescope.
The magnitude of ground motion at the site is commonly related to the occurrence rate of scattered light noise \cite{Accadia:2010zzb, Longo:2022, Soni:2021, Soni:2025}.
In Virgo, the scattered light noise primarily originated from the back-scattering of the optical components located on external benches outside the vacuum system \cite{Accadia:2010zzb}.
In LIGO, it originated from the phase-modulated back-scattering of the electrostatic drive on the annular end reaction mass, caused by micron-scale relative motion with the end test mass \cite{Soni:2021}.

Research from Virgo and LIGO utilized methods based on empirical mode decomposition and the Hilbert-Huang transform to identify the periods of the scattered light noise in the data \cite{Longo:2022, Guillermo:2017}.
These methods provide time-varying frequency features in the data.
By computing the correlation between the decomposed frequency feature and the estimated noise from motion of known sources, they recovered the scattered light noise in the data  \cite{Longo:2022}.
Since these methods require the identification of the scattering object in advance, we proposed a new algorithm, named f-cluster, that inputs only the time-frequency maps of the data.

%%% comparasion of old & new method
The f-cluster algorithm characterizes the periodicity of the excess power regions in a window of a time-frequency map.
Based on the periodicity, the algorithm can detect periodic arch-shaped features of the scattered light noise.
The f-cluster algorithm has two advantages. Firstly, unlike the previous methods \cite{Longo:2022, Guillermo:2017}, which require the velocity of the identified scattering object to be measured by other sensors, the f-cluster algorithm is agnostic to the scattering object.
Therefore, our algorithm can be used for noise hunting to identify the origin of the scattering object.

Secondly, the f-cluster algorithm can provide the occurrence time and the precise frequency of the scattering object, whereas the previous methods can only provide frequency bands, including scattered light noise.
As a result, we can quantitatively investigate the relationship between the scattered light noise occurrence rate and the magnitude of seismic motion.
In this paper, we applied our algorithm to the KAGRA data from April 2025 as a demonstration.

% 5. Overview of the following sections.
In Section 2, we introduce the behavior of scattered light noise in interferometers.
In Section 3, we present a detailed procedure for the newly developed f-cluster algorithm.
In Section 4, we demonstrate the performance of the f-cluster algorithm using KAGRA data.

\section{Scattered Light Noise}

\subsection{Introduction of the scattered light noise}
The scattered light noise is one of the non-stationary noises caused by recombining the diffused light into the main optical path. 
In past research \cite{Accadia:2010zzb}, a model of scattered light noise in the interferometer has been proposed.
The noise introduced by the scattered light $h_{sc}(t)$ can be written as
\begin{equation} \label{eq:h}
    h_{sc}(t) = G\cdot\sin{\left(\frac{4\pi}{\lambda}(x_0 + \delta x_{sc}(t))\right)},
\end{equation}
where $G$ is the coupling factor, $\lambda$ is the wavelength of the laser, $x_0$ is the static light path, and $\delta x_{sc}(t)$ is the displacement of the diffused light along the direction of the static light path. 
Due to the time variation of $\delta x_{sc}(t)$, the characteristic frequency of the scattered light noise also has time variation, which can be written as
\begin{equation} \label{eq:farch}
    f_\text{arch}(t) = \frac{1}{2\pi}\left| \frac{d}{dt}\left(\frac{4\pi}{\lambda}(x_0 + \delta x_{sc}(t))\right) \right| = 2\left| \frac{v_{sc}(t)}{\lambda} \right|,
\end{equation}
where $v_{sc}(t)$ is the velocity of the scattering object.
Eq. \ref{eq:farch} demonstrates the characteristic frequency due to a single round-trip of the diffused light.
If there are $N$ round-trips, the characteristic frequency of the scattered light noise is modified as
\begin{equation} 
    f_{\text{arch},N}(t) = N \cdot f_\text{arch}(t) = N \cdot 2\left| \frac{v_{sc}(t)}{\lambda} \right|.
    \label{eq:farchN}
\end{equation}

\begin{figure}[H]
    \centering
    \includegraphics[width=0.7\linewidth]{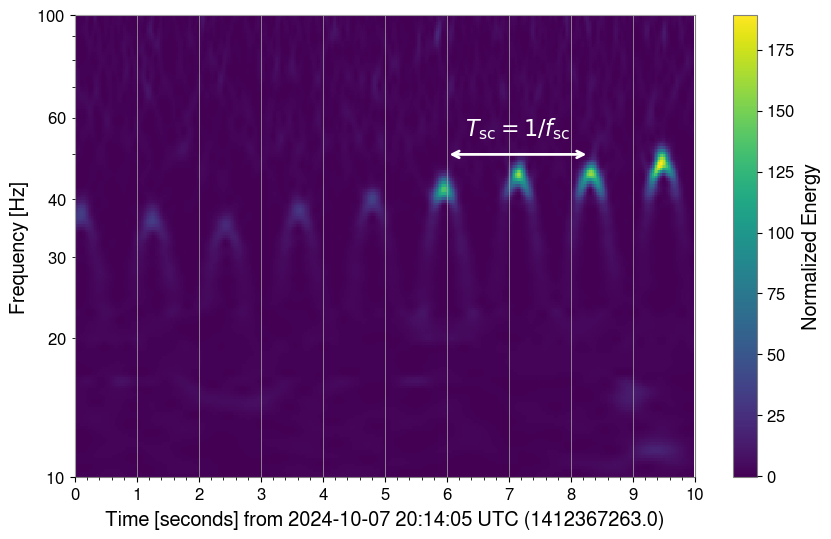}
    \caption{
    An example of the scattered light noise on a time-frequency map.
    The data is taken in the STRAIN channel in the KAGRA data.
    Twice the time difference between the neighboring arch shapes refers to one period of the motion of the scattering source, with a period of $T_{sc} = 2.3$ s in KAGRA.
    }
    \label{fig:SL_example}
\end{figure}

This model predicts the appearance of a characteristic arch-shaped pattern in time-frequency maps as $f_\text{arch}(t)$. 
Figure \ref{fig:SL_example} shows the characteristic arch shapes of the scattered light noise observed in KAGRA data.
To connect the feature of scattered light noise with the scattering object motion, we can use Eq. \ref{eq:farch}.
Suppose the scattering motion $x_{sc}(t)$ is simple harmonic with period $T_{sc} = 1/f_{sc}$; its velocity $v_{sc}(t)$ is also simple harmonic.
Hence, $f_\text{arch}(t)$ is periodic with period $\frac{1}{2}T_{sc}$.
That is, within one period of the scattering motion, two scattered light arch shapes appear in a time-frequency map.

The scattered light noise contaminates the sensitivity of KAGRA in the $30-100$ Hz band, which is the most sensitive frequency band for compact binary coalescence searches. 
We observe the scattered light noise in the degrees of freedom of the differential arm length in the interferometer.
The precise source of the scattered light noise in KAGRA has not yet been identified.
We found that the periodicity of scattered light noise in KAGRA is stable at $T_{sc}\approx 2.3$ s or $f_{sc} \approx 0.43$ Hz.
The magnitude of other noise sources limiting KAGRA sensitivity is currently more significant than the scattered light noise.
Given that scattered light noise has been a barrier to improving the sensitivity of overseas observatories, it is anticipated that this noise will also limit sensitivity at KAGRA in the future.
To identify the cause and improve the situation, it is essential to determine the occurrence time and frequency using systematic methods.

This work aims to characterize the scattered light noise in the KAGRA interferometer and its relation to the seismic motions at the KAGRA site.
With the knowledge of the relation between the scattered light noise and seismic motion, we can further forecast the scattered light noise in the interferometer by a threshold of the magnitude of seismic motion.
One of the challenges in analyzing the scattered light noise is the nonlinear coupling between the source and the signal.
Since the critical cause of the scattering source of the scattered light noise in KAGRA is unclear, we developed an algorithm to identify the scattered light noise in time-frequency maps and analyze its features within the interferometer.
This algorithm focuses on the excess power in the time-frequency maps, also known as the spectrogram.
In this analysis, the input to the algorithm is a 30-second constant Q-transformed spectrogram \cite{Brown:1991}.
The algorithm outputs the periodicity of the regions with excess power.
Eq. \ref{eq:farch} describes the excess of power regions in time-frequency maps for scattered light noise.
Thus, by supposing the relative motion between the mirrors and the scattering surface to be simple harmonic, we can provide constraints on the periodicity to target the scattered light noise. 

\subsection{KAGRA Data}
In KAGRA data, the characteristic arch-shaped feature from scattered light noise was observed not only in the calibrated DARM (differential arm length) data, known as STRAIN, but also in the other length degrees of freedom data, such as MICH (short Michelson interferometer) and PRCL (power recycling cavity length).
The KAGRA interferometer is a power-recycled Fabry-Perot-Michelson interferometer.
The interferometer consists of multiple optical cavities that require control, and thus, the cavities are in resonance with the laser.
The detailed definition of the mirrors, pick-up ports, and length degrees of freedom is introduced in \cite{kagra:2023} Section 2.

In this analysis, we focused on periods when the interferometer was in the locked state.
When the interferometer is locked, all cavities are maintained in resonance, producing a dark fringe at the antisymmetric port \cite{kagra:2023}.
The total duration of locked data used for this analysis in April 2025 is 273 hours.

%%%%%%%%%%%%%%%%%%%%%%%%
\section{Methodology}
In this section, we present the detailed procedure for the methods that identify and characterize scattered light noise in KAGRA data.
We developed the f-cluster algorithm, designed to analyze the periodicity of excess power regions in the spectrograms.
The essential method for characterizing scattered light noise is the constant-Q transformation and the f-cluster algorithm.
The constant Q-transform provides spectrograms with a designed frequency resolution to represent the typical arch shapes of scattered light noise.
The f-cluster algorithm enables us to detect scattered light noise in the data and characterize the periodicity of its arch-shaped features, thereby determining the frequency of the scattering motion from the source.

\subsection{Constant Q-Transform}
\begin{figure}[H]
    \centering
    \includegraphics[width=1\linewidth]{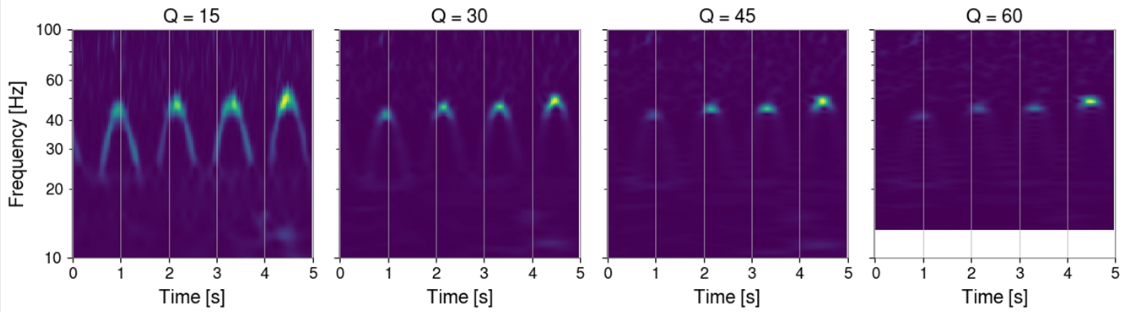}
    \caption{
    Example of Q-transformed spectrogram of the scattered light noise with different Q values, $Q = 15, 30, 45, 60$, which is computed from the KAGRA data since 2024-10-17 20:14:00 UTC.
    }
    \label{fig:qcompare}
\end{figure}

The Q-transform is a modification of the short-time Fourier transform.
The analysis window duration varies inversely with frequency such that tiles of constant quality factor $Q$ cover the time-frequency spectrogram. 
As a result, the frequency resolution around frequency $f$ is proportional to $f/Q$ and the time resolution is proportional to $Q/f$.
A formula of the constant Q-transform \cite{Brown:1991,Robinet:2020} for a time series $x(t)$ is
\begin{equation}
    X(t,f,Q) = \int_{-\infty}^\infty x(\tau)w(\tau-t,f,Q)e^{-2i\pi f \tau}\; d\tau.
    \label{eq:Q_eq}
\end{equation}
In Eq. \ref{eq:Q_eq}, $w(\tau-t,f,q)$ is a Gaussian window function defined as
\begin{equation}
    w(\tau-t,f,Q) = \frac{W_g}{\sigma_t\sqrt{2\pi}}\exp\left( -\frac{1}{2\sigma_t^2}(\tau-t)^2 \right),
\end{equation}
where $W_g$ is the normalization constant, and $\sigma_t$ is the Gaussian variance defined by
\begin{equation}
    \sigma_t^2 = \frac{Q^2}{8\pi^2f^2}.
\end{equation}
The Q-transforms with different Q values offer varying resolutions of signals in both time and frequency.
Figure \ref{fig:qcompare} is an example of the Q-transformed spectrograms of the scattered light noise for $Q = 15, 30, 45, 60$.
Hence, we selected $Q = 30$ to take a balance between frequency and time resolution, representing the targeted arch shapes.
In the analysis, we applied the Q-transform methods from GWpy \cite{gwpy:2021}.

\subsection{f-cluster Algorithm}
\label{sec:f_cl}
A feature of the scattered light noise is the arch-shaped patterns in the spectrograms, which result in the excess power regions.
As shown in figure \ref{fig:SL_example}, the spectrogram can visualize the characteristic arch-shaped pattern of the scattered light noise.
Exploiting this characteristic, we developed an algorithm to detect regions of excess power and quantify their periodicity.
From the time-series data, we generated a spectrogram, extracted the regions of excess power, performed clustering, and determined the periodicity of the clusters.
Because visual identification by manual inspection is inefficient and subjective, an automated algorithm is required to reliably identify and characterize scattered light noise.
The proposed method significantly reduces reliance on manual analysis, improving the efficiency and reproducibility of the identification process.

The f-cluster algorithm analyzes the 30-second Q-transformed spectrogram and outputs the period $T_\text{cluster}$, its standard deviation $\sigma_T$, the frequency $f_{cluster}$, and its standard deviation $\sigma_f$, which represent the periodicity of the clusters.
Figure \ref{fig:flow} shows a flowchart of the f-cluster algorithm.
Suppose we apply the algorithm to the data starting from $t_0$ to $t_0 + 30$ s.
The procedure of the analysis is as follows:

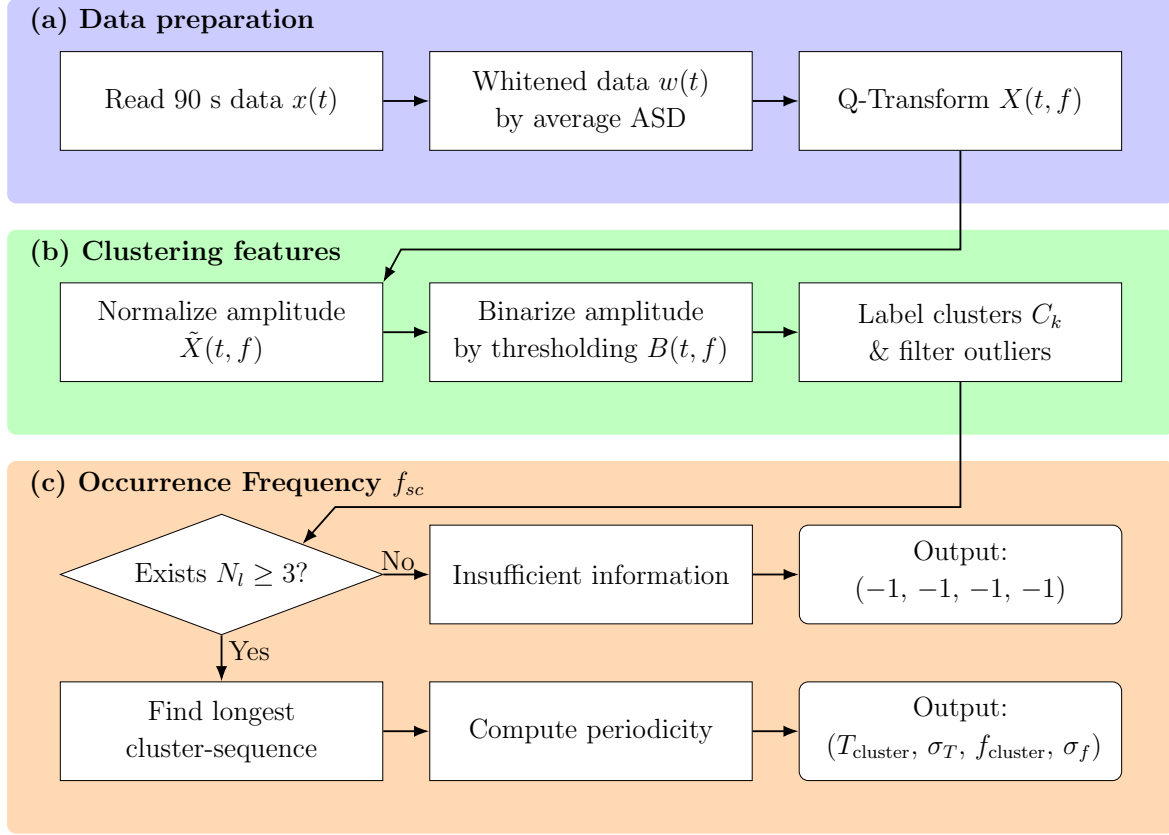
\begin{figure}[H]
    \centering
    % \usepackage{tikz}
% \usetikzlibrary{arrows.meta, positioning, shapes.geometric, fit, backgrounds}

% Global box size (fixed for all rectangular nodes)
\newlength{\boxw}\setlength{\boxw}{45mm} % width
\newlength{\boxh}\setlength{\boxh}{15mm} % height

\resizebox{\textwidth}{!}{
\begin{tikzpicture}[
  node distance = 7mm and 7mm,
  >=Latex,
  line/.style={-Latex,thick},
  block/.style={
    draw, fill=white,align=center,
    text width=\boxw, minimum height=\boxh, inner sep=2mm
  },
  output/.style={
    draw, rounded corners,fill=white,align=center,
    text width=\boxw, minimum height=\boxh, inner sep=2mm
  },
  decision/.style={
    diamond,draw,align=center,fill=white,aspect=2,
    inner sep=1.5pt,minimum width=49.05mm, minimum height=\boxh
  },
  title/.style={font=\bfseries,anchor=north west}
]

    % ---------- (a) Data preparation ----------
    \node[block]                 (a1) {Read 90 s data $x(t)$};
    \node[block,right=of a1]     (a2) {Whitened data $w(t)$ \\ by average ASD};
    \node[block,right=of a2]     (a3) {Q-Transform $X(t,f)$};
    
    \draw[line] (a1) -- (a2);
    \draw[line] (a2) -- (a3);
    
    % ---------- (b) Labeling arches ----------
    \node[block,below=20mm of a1] (b1) {Normalize amplitude \\ $\tilde{X}(t,f)$};
    \node[block,right=of b1]      (b2) {Binarize amplitude \\ by thresholding $B(t,f)$};
    \node[block,right=of b2]      (b3) {Label clusters $C_k$ \\ \& filter outliers};
        
    \draw[line] (a3.south) -- ++(0,-15mm) -- ++(-85mm,0) -- (b1.north east);
    \draw[line] (b1) -- (b2);
    \draw[line] (b2) -- (b3);
    
    % ---------- (c) Occurrence Frequency ----------
    \node[decision,below=20mm of b1] (c0) {Exists \(N_l \ge 3\)?};
    
    \node[block,below=of c0] (c1) {Find longest \\ cluster-sequence};
    \node[block,right=of c1]      (c2) {Compute periodicity};
    \node[output,right=of c2]     (c3)
    {Output:\\\begin{tabular}{c}
    \((T_{\text{cluster}},\, \sigma_T,\) \(f_{\text{cluster}},\, \sigma_f)\)
    \end{tabular}};
    
    \node[block,right=of c0]      (c4) {Insufficient information};
    \node[output,right=of c4]     (c5)
    {Output:\\\begin{tabular}{c}
    \((-1,\,-1,\) \(-1,\,-1)\)
    \end{tabular}};
    
    \draw[line] (b3.south) -- ++(0, -19mm) -- ++(-95mm,0) -- (c0.north east);
    
    \draw[line] (c0.south) -- node[above,pos=0.75, xshift=12pt]{Yes} (c1.north);
    \draw[line] (c1) -- (c2);
    \draw[line] (c2) -- (c3);
    
    \draw[line] (c0.east) -- node[left,pos=0.9,yshift=6pt]{No} (c4.west);
    \draw[line] (c4) -- (c5);
    
    % ---------- Colored bands behind (fit) ----------
    \begin{scope}[on background layer]
      \node[rounded corners,fill=blue!20,inner sep=8mm,fit=(a1)(a2)(a3)] (A) {};
      \node[rounded corners,fill=green!25,inner sep=8mm,fit=(b1)(b2)(b3)] (B) {};
      \node[rounded corners,fill=orange!30,inner sep=8mm,fit=(c0)(c1)(c2)(c3)(c4)(c5)] (C) {};
    \end{scope}
    
    % ---------- Section titles ----------
    \node[title] at ([xshift=2mm,yshift=-0mm]A.north west) {(a) Data preparation};
    \node[title] at ([xshift=2mm,yshift=-0mm]B.north west) {(b) Clustering features};
    \node[title] at ([xshift=2mm,yshift=-0mm]C.north west) {(c) Occurrence Frequency \(f_{sc}\)};

\end{tikzpicture}
}   
    \caption{
    The flowchart of the f-cluster algorithm to characterize the feature of excess power regions in spectrograms:
    (a) From the time series data $x(t)$, we prepare spectrograms $X(t,f)$ by whitening and the  Q-transform,
    (b) We identify and cluster the features from $X(t,f)$ by thresholding, binarizing, and grouping the excess power regions, which we refer to as the clusters.
    (c) We assign the clusters to different event sequences.
    In the case that there is a sequence with length $N_l$ that is larger than or equal to 3, the algorithm outputs the periodicity of the longest sequence by $(T_\text{cluster}, \sigma_T, f_\text{cluster}, \sigma_f)$.
    Otherwise, we set the output $(T_\text{cluster}, \sigma_T, f_\text{cluster}, \sigma_f) = (-1,-1,-1,-1)$ indicating insufficient information to compute the periodicity.
    }
    \label{fig:flow}
\end{figure}

\renewcommand{\thesubsubsection}{(\alph{subsubsection})}
\subsubsection{Data preparation} 
\begin{figure}[H]
    \centering
    \includegraphics[width=0.8\linewidth]{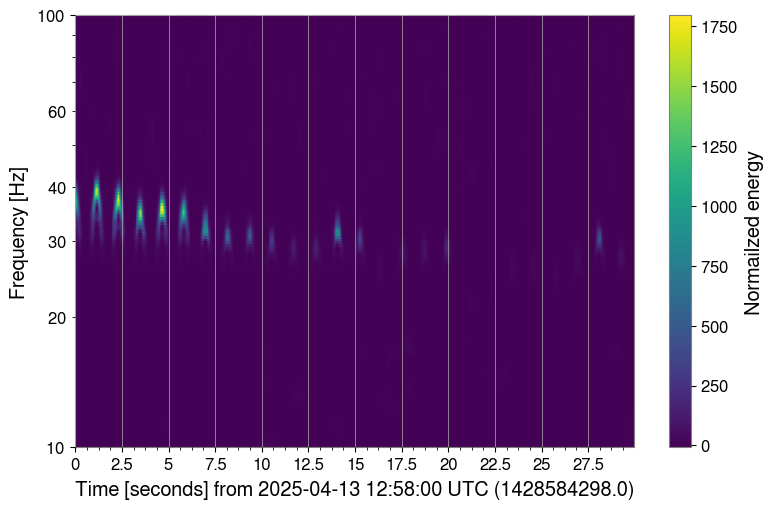}
    \caption{
      The example of a Q-transformed spectrogram of the whitened data $X(t,f)$ includes several arch-shaped regions corresponding to the scattered light noise.
      }
    \label{fig:flow_org}
\end{figure}

\hfill\\
First, we read $90$ seconds of time series data $x(t)$ from $t_0-30$ to $t_0+60$ s. 
If this region overlaps with the time when the interferometer is not locked, then we read $t_0-60$ to $ t_0+30$ s or $t_0$ to $ t_0+90$ s.
Then, we estimate the amplitude spectrum density (ASD) using the median method on the 90-second data, with a Fourier transform length of four seconds and an overlap of two seconds between adjacent segments.
We design the whitening filter based on the estimated ASD, and whiten $x(t)$ to the whitened data $w(t)$ by using convolution with the whitening filter.
After that, we crop the edges of the whitened time series $w(t)$ such that $w(t)$ is defined in $t_0-3$ to $t_0+33$ s.
We perform a Q-transform on the whitened data and crop the Q-transformed spectrogram in time from $t_0$ to $t_0+30$ s to avoid edge effects caused by applying the time-domain filter. 
Also, we crop the frequencies from 10 to 100 Hz to focus on the contaminated band. 
Figure \ref{fig:flow_org} shows the example of the obtained spectrogram $X(t, f)$.

%%%%%%%%%%%%%%%%%%%%%%%%%%%%%%%%%%%%%%%%%%%%%%%%%%%%%%
\subsubsection{Clustering features} \hfill\\
After obtaining the Q-transformed spectrogram, if scattered light noise is present, we aim to detect it and calculate the periodicity of its characteristic arch shapes.
Therefore, a process of labeling the characteristic arch shapes is necessary.
In our methodology, each of these arch shapes is treated as a single cluster — a distinct group of neighboring excess power pixels in the time-frequency map.

\begin{figure}[H]
    \centering
    \includegraphics[width=0.8\linewidth]{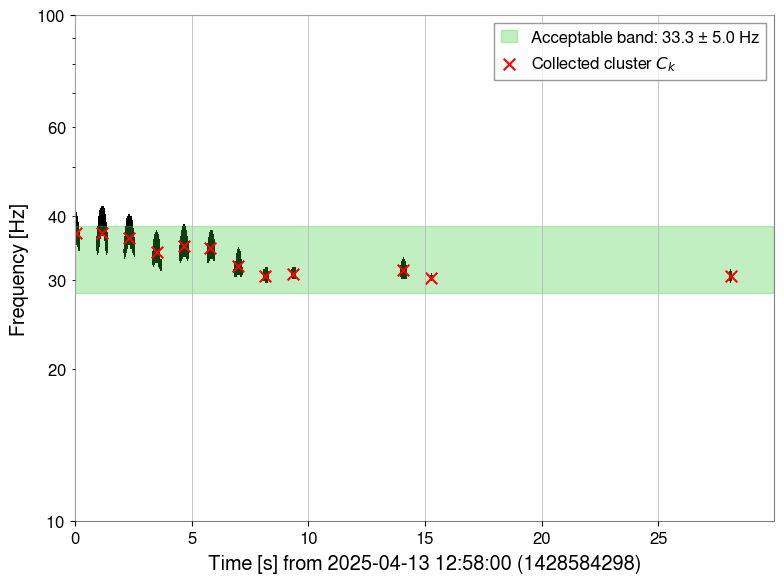}
    \caption{
      The example of binary spectrogram $B(t,f)$, which is converted from the normalized Q-transformed spectrogram $\tilde{X}(t,f)$ by using Eq. \ref{eq:bimap}.
      The black pixels stand for $B(t,f)=1$, while the white pixels represent $B(t,f) = 0$.
      In this case, $\bar{f}_\text{inc} = 33.3$ Hz and $\sigma_\text{inc} = 2.6$ Hz.
      The green band indicates the acceptable frequency band, which is $33.3 \pm 5.0$ Hz, and the red crosses indicate the cluster satisfying the condition Eq. \ref{eq:inc}.
      }
    \label{fig:flow_bi}
\end{figure}

We normalize the spectrogram $X(t,f)$ to the normalized spectrogram $\tilde{X}(t,f)$, which is defined as
\begin{equation}
    \tilde{X}(t,f) = \frac{X(t,f) - \min(X)}{\max(X) - \min(X)},
\end{equation}
where functions $\max(X)$ and $\min(X)$ take the maximum or the minimum value from $X(t,f)$.
To obtain the position of arch shapes from scattered light noise or the other regions in a spectrogram, which corresponds with the regions of excess power, we apply a threshold to the normalized spectrogram $\tilde{X}(t,f)$.
We sort all the pixels of $\tilde{X}(t,f)$ from small to large, and take the 99.5th percentile of the sorted $\tilde{X}(t,f)$ as the threshold $p$.
Then, we define a binary mapping to map the normalized spectrogram $\tilde{X}(t,f)$ to a binary spectrogram $B(t,f)$ by

\begin{equation}
    B(t,f) = \theta( \tilde{X}(t,f) - p ),
    \label{eq:bimap}
\end{equation}
where $\theta(x-p)$ is the unit step function defined as
\begin{equation}
    \theta(x-p) = \begin{cases}
        0, & x < p \\
        1, & x \geq p
    \end{cases}\; .
\end{equation}
Figure \ref{fig:flow_bi} shows the example of $B(t,f)$.
There are several groups of black pixels of $B(t,f)=1$ that satisfy $\tilde{X}(t,f) \geq p$.
The other pixels in $B(t,f)$ are considered as the background and set to zero.

In figure \ref{fig:flow_bi}, the characteristic arch shapes of scattered light noise are mapped to one as $B(t,f)=1$, which consists of the black pixels.
To analyze these features, we now transition from treating the spectrogram as a continuous map to analyzing the discrete set of activated pixels.
We cluster the pixels of neighboring ones by depth-first search \cite{Tarjan:1972}, and denote the $k$-th cluster of pixels as a set $A_k$ of pixel positions $(t_{k, i}, f_{k,j})$.
We define the set of pixels $A_k$ as
\begin{equation}
    A_k = \{ (t_{k, i}, f_{k,j})\; |\; B(t_{k, i}, f_{k,j}) = B(t_{k, i\pm1}, f_{k,j\pm1}) = 1 \},
    \label{eq:A_k}
\end{equation}
where $(t_{k, i}, f_{k,j})$ stands for the pixel at the $i$-th column and the $j$-th row in a spectrogram.
The centroid of the cluster $C_k$ is the representation of the $k$-th cluster $A_k$
\begin{equation}
    C_k = (t_k, f_k) = \frac{1}{n_k}\sum_{(t_{k, i}, f_{k,j}) \in A_k}(t_{k, i}, f_{k,j}),
    \label{eq:C_k}
\end{equation}
where $n_k$ denotes the number of pixels in $k$-th cluster.
The procedures for obtaining the clusters $C_k$ refer to the clustering process.

After clustering, we apply a filter to exclude the outliers.
Since our analysis target is the scattered light noise, this filter is based on the continuous nature of the scattered light noise.
Since $f_\text{arch}(t)$ from Eq. \ref{eq:farch} is continuous in time, if there is scattered light noise, we expect there are multiple clusters $C_k = (t_k, f_k)$ with their vertical positions $f_k$ to be distributed within a finite frequency band.
That is, if there are outlier clusters from other kinds of noise located in a different frequency band, we want to exclude them first.
Hence, we establish the following condition of the acceptable frequency band with regard to the average $\bar{f}_\text{inc}$ and the standard deviation $\sigma_\text{inc}$ of the vertical positions $f_k$ for all $C_k$ in a spectrogram.
\begin{equation}
    |f_k - \bar{f}_\text{inc}| \leq \max(s\cdot\sigma_\text{inc}, 5\; \text{Hz}),
    \label{eq:inc}
\end{equation}
where the parameter $s$ controls the acceptable frequency band width in units of $\sigma_\text{inc}$.

We tuned $s=1.5$ empirically to keep in-band clusters and reject distant outliers.
The $\max(s\cdot\sigma_\text{inc}, 5)$ function introduces a robust minimum acceptable width of $\pm 5$ Hz even when the initial cluster set has a near-zero variance, preventing the acceptable band from becoming overly restrictive.
We only collect the clusters $C_k = (t_k, f_k)$ satisfying the condition of Eq. \ref{eq:inc} for the next analysis step.

%%%%%%%%%%%%%%%%%%%%%%%%%%%%%%%%%%%%%%%%%%%%%%%%%%%%%%
\subsubsection{Occurrence frequency} \hfill\\
After labeling the regions of excess power as the clusters $C_k$, we compute the periodicity of the clusters.

With the clusters $C_k$ obtained from $B(t,f)$ by Eq. \ref{eq:A_k} and \ref{eq:C_k}, we introduce a new property called \textit{consecutive} by the time difference between the cluster $C_k$ and the next cluster $C_{k+1}$, whose clusters have a centroid position of $(t_k, f_k)$ and $(t_{k+1}, f_{k+1})$, respectively.
If the time difference between the two clusters is less than an upper bound $\Delta T$, that is
\begin{equation}
    t_{k+1} - t_k < \Delta T,
    \label{eq:csct}
\end{equation}
then clusters $C_k = (t_k, f_k)$ and $C_{k+1}$ are consecutive.
By applying the condition of Eq. \ref{eq:csct} to the clusters, we construct the sequence $S_l$, defined by 
\begin{equation}
    S_l = \left\{ C_k, C_{k+1} |\; t_{k+1} - t_k < \Delta T \right\},
\end{equation}
where we set $\Delta T = 1.5$ s because the time difference between two arches of the scattered light noise observed in KAGRA is around 1.2 s, and an extra 0.3 s was added.
As a result, multiple sequences can exist in a spectrogram due to the distribution of the clusters in time.

Let the number of clusters in a sequence $S_l$ be $N_l$.
For all sequences in a spectrogram, the procedures of the algorithm are divided into two cases depending on the $N_l$ of the sequences.
In the case that there exists any sequence with $N_l \geq 3$, we find the longest sequence, and compute the average period and average frequency of the clusters in this sequence $(T_\text{cluster}, f_\text{cluster})$ and their standard deviation $(\sigma_T, \sigma_f)$, as the output of the algorithm:
\begin{equation}
    \begin{cases}
        T_\text{cluster} &= \frac{1}{N_l -1}\sum_{i=1}^{N_l-1}\;T_i\\
        \sigma_T &= \sqrt{\frac{1}{N_l -1}\sum_{i=1}^{N_l-1}(T_i - T_\text{cluster})^2}
    \end{cases},
\end{equation}
and 
\begin{equation}
    \begin{cases}
        f_\text{cluster} &= \frac{1}{N_l -1}\sum_{i=1}^{N_l-1}\;f_i\\
        \sigma_f &= \sqrt{\frac{1}{N_l -1}\sum_{i=1}^{N_l-1}(f_i - f_\text{cluster})^2}
    \end{cases},
    \label{eq:f_cluster}
\end{equation}
where $T_i = t_{i+1} - t_i$ and $f_i = 1/T_i$ for $i=1 \cdots N_l-1$. \footnote{
We can relax the condition to $N_l \geq 2$. 
If $N_l = 2$ for a sequence, $T_{cluster} = T_1, f_{cluster} = f_1$, and their standard deviation $\sigma_T, \sigma_f$ do not provide information about the periodicity of the consecutive clusters in the sequence. 
With the relaxed condition $N_l \ge 2$, more possible scattered light noise can be labeled, yet it yields more false alarms than using the condition $N_l \ge 3$. 
Therefore, we use $N_l \ge 3$ as the condition in the f-cluster algorithm.
}
In the case of no sequence with $N_l \geq 3$, the information is considered insufficient to compute the periodicity of the clusters in the spectrogram.
Thus, we set the output to be $(T_\text{cluster}, \sigma_T, f_\text{cluster}, \sigma_f)=(-1, -1, -1, -1)$ as the drop-out.

From the nature of the scattered light noise described by Eq. \ref{eq:farch}, $f_\text{arch}(t) = 2|\frac{v_{sc}(t)}{\lambda}|$, when the motion of the scattering object being approximated by a simple harmonic motion with period $T_{sc} = 1/f_{sc}$, the period of $f_\text{arch}(t)$ is $\frac{1}{2}T_{sc}$.
In the f-cluster algorithm, we compute the periodicity of the excess power regions in a spectrogram as $(T_\text{cluster}, \sigma_T, f_\text{cluster}, \sigma_f)$.
If the scattered light noise is observed in the spectrogram, the periodicity of the scattering motion and the excess-power regions are linked to the occurrence frequency $f_{sc}$ and its standard deviation $\sigma_{sc}$ as follows:
\begin{equation}
    \begin{cases}
        f_{sc} &= \frac{1}{2}f_\text{cluster} \\
        \sigma_{sc} &= \frac{1}{2}\sigma_f
    \end{cases}.
    \label{eq:fsc}
\end{equation}

\begin{figure}[H]
    \centering
    \includegraphics[width=0.8\linewidth]{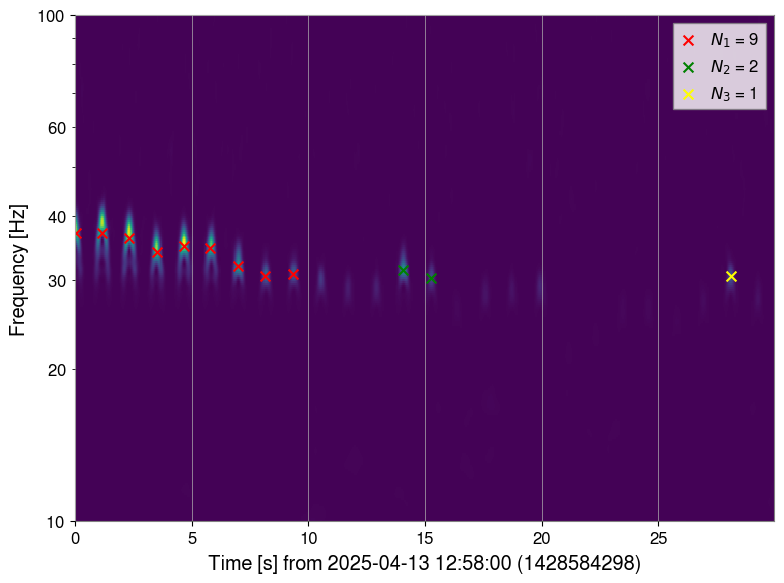}
    \caption{
      An example of consecutive clusters in the presence of scattered light noise in MICH data.
      The red crosses indicate the arches clustered from $B(t, f)$, which is the longest consecutive cluster sequence.
      The green and yellow crosses indicate the other two sequences whose number of clusters is not the longest.
      The consecutive cluster sequence satisfies the condition of Eq. \ref{eq:csct}.
      }
    \label{fig:flow_xxx}
\end{figure}

Figure \ref{fig:flow_xxx} illustrates an example of consecutive clusters in the presence of scattered light noise.
The red crosses indicate the arches clustered from $B(t, f)$, with the longest consecutive cluster sequence with $N_1 = 9$.
For this sequence, by using Eq. \ref{eq:f_cluster} and Eq. \ref{eq:fsc}, we computed the occurrence frequency $f_{sc} = 0.431$ Hz and the standard deviation $\sigma_{sc} = 0.011$ Hz.

\subsection{Method validation}
\begin{figure}[H]
    \centering
    \begin{subfigure}[h]{1\textwidth}
    \includegraphics[width=1\linewidth]{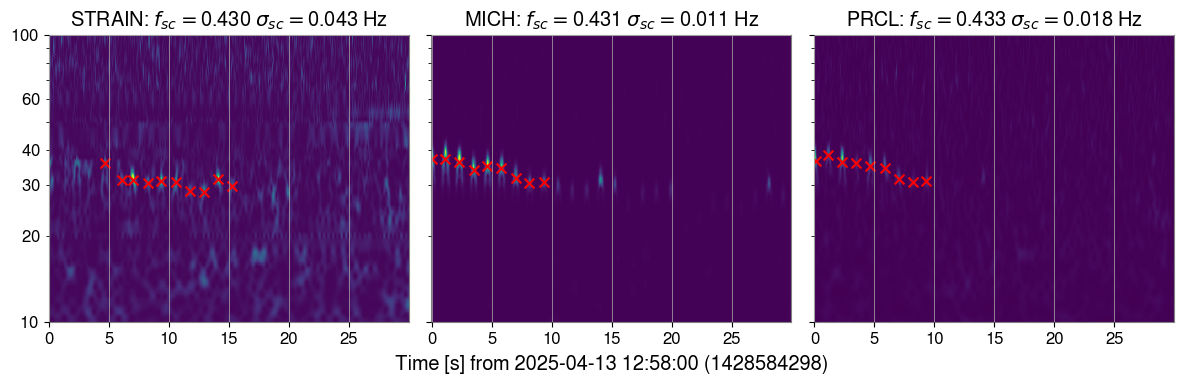}
    \caption{
    An example of successfully detecting the scattered light noise in three channels.
    The occurrence frequency is estimated to be  $f_{sc} = 0.431$ Hz with the standard deviation $\sigma_{sc}<0.05$ Hz.
    }
    \label{fig:ex_a}
    \end{subfigure}

    \begin{subfigure}[h]{1\textwidth}
    \includegraphics[width=1\linewidth]{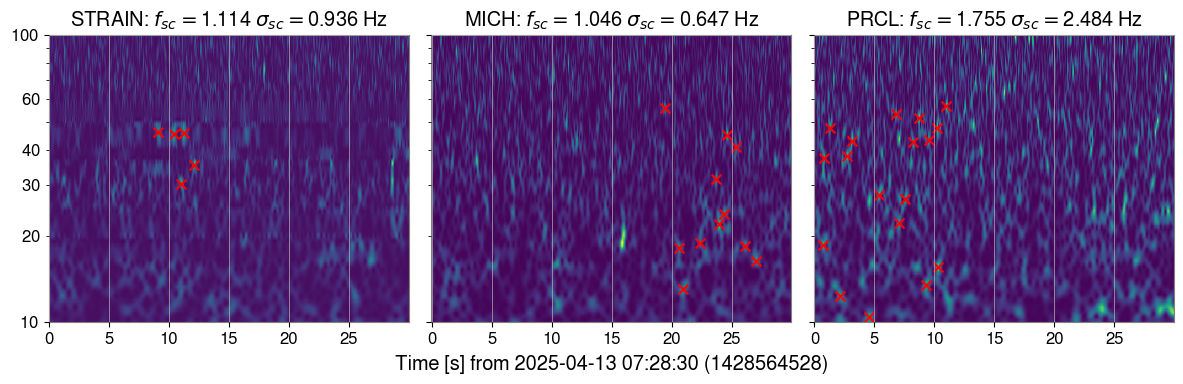}
    \caption{
    An example of applying the f-cluster algorithm to the background.
    In three channels, the standard deviation $\sigma_{sc} > 0.5$ Hz, which indicates randomness.
    }
    \label{fig:ex_c}
    \end{subfigure}
    \caption{
    The examples demonstrate the f-cluster algorithm applied to KAGRA data in the STRAIN (left), MICH (middle), and PRCL (right) channels.
    Figure \ref{fig:ex_a} contains the scattered light noise, while the figure \ref{fig:ex_c} contains no common patterns.
    The title labels the channel, its occurrence frequency $f_{sc}$, and standard deviation $\sigma_{sc}$.
    The red crosses only denote the position of the longest consecutive sequence.
    }
    \label{fig:algo_exmp}
\end{figure}

As described in Section \ref{sec:f_cl}, the f-cluster algorithm provides the periodicity of the excess regions on a Q-transformed spectrogram by the period, frequency, and their standard deviation as $(T_\text{cluster}, \sigma_T, f_\text{cluster}, \sigma_f )$.
For the scattered light noise, by Eq. \ref{eq:fsc}, we introduced the occurrence frequency $f_{sc} = \frac{1}{2} f_\text{cluster}$ to characterize the frequency of the motion of the scattering object.
Figure \ref{fig:algo_exmp} shows examples of performing the f-cluster algorithm on the KAGRA data.
Figures \ref{fig:ex_a} show the spectrograms of the scattered light noise and the characteristic arch shapes of the scattered light noise.
From the output of the algorithm, the occurrence frequency is estimated to be $f_{sc} = 0.431$ Hz, and the standard deviation $\sigma_{sc}$ in three channels is less than 0.05 Hz.
Figures \ref{fig:ex_c} are examples of analyzing the data without the scattered light noise.
Since there is no special pattern in the spectrogram, the standard deviation $\sigma_{sc}$ is larger, which indicates the randomness.

Therefore, we apply the following conditions to $\sigma_T$ and $f_{sc}$, to separate the scattered light noise and background noise,
\begin{equation}
    \begin{cases}
        \sigma_T < 0.3\; \text{s}\\
        0.4 \leq f_{sc} \leq 0.5 \; \text{Hz}.
    \end{cases}
    \label{eq:condition}
\end{equation}
The first condition is to highlight the periodic occurrence of scattered light noise, where 0.3 s is about half the width of a single arch shape.
The second condition focuses on the scattered light noise that is currently of interest. 
If the $f_{sc}$ and $\sigma_T$ computed from the 30-second spectrogram meet the above conditions, we assign an SL (scattered light) label to the 30-second time span.

\begin{figure}[H]
    \centering
    \begin{subfigure}[h]{0.49\textwidth}
      \includegraphics[width=1\linewidth]{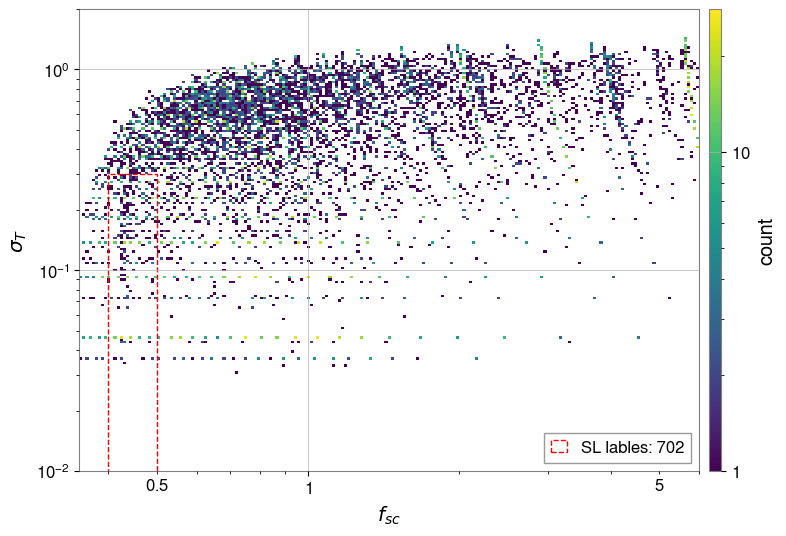}
      \caption{STRAIN} \label{fig:6a}
    \end{subfigure}
    \hspace*{\fill}   % maximize separation between the subfigures
    \begin{subfigure}[h]{0.49\textwidth}
      \includegraphics[width=1\linewidth]{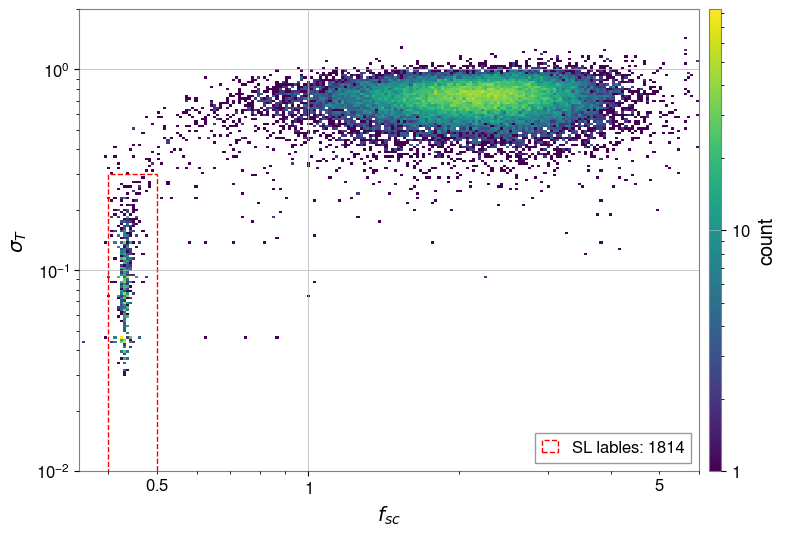}
      \caption{MICH} \label{fig:6b}
    \end{subfigure}
    \hspace*{\fill}   % maximize separation between the subfigures
    \begin{subfigure}[h]{0.49\textwidth}
      \includegraphics[width=1\linewidth]{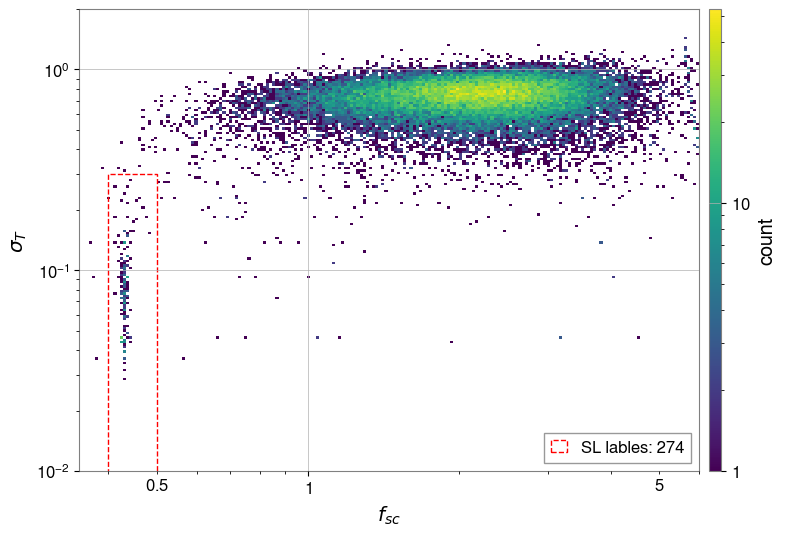}
      \caption{PRCL} \label{fig:6c}
    \end{subfigure}
    \hspace*{\fill}   % maximize separation between the subfigures
    \caption{
    Two-dimensional histogram of $f_{sc}$ and $\sigma_T$ obtained from the f-cluster algorithm to STRAIN, MICH, and PRCL channels. \\
    STRAIN: number of valid $f_{sc}$ outputs: 11963; SL labels: 702; $f_{sc} = 0.445 \pm 0.027$ Hz \\
    MICH: number of valid $f_{sc}$ outputs: 30558; SL labels: 1814; $f_{sc} = 0.432 \pm 0.009$ Hz \\
    PRCL: number of valid $f_{sc}$ outputs: 32018; SL labels: 274 ; $f_{sc} = 0.431 \pm 0.010$ Hz\\
    The data is collected in the KAGRA interferometer during April 2025.
    Note that, due to the noise background in STRAIN is more complex than in MICH or PRCL data, the input spectrogram of STRAIN data is cropped to $20-40$ Hz.
    }
    \label{fig:algo_vali}
\end{figure}

Figure \ref{fig:algo_vali} shows the two-dimensional histogram of $f_{sc}$ and $\sigma_T$ from April 1, 2025, to April 30, 2025. 
The red dashed box represents the region of the SL label condition in Eq. \ref{eq:condition}. 
\noindent
Figure \ref{fig:algo_vali} shows that the algorithm outputs form two primary populations, which are well-separated by the periodicity $\sigma_T$.
The first population is a dense concentration localized within the red dashed box, satisfying the SL label constraints.
This group's periodicity standard deviation ($\sigma_T < 0.3$ s) indicates a stable, periodic occurrence of the excess power regions, and its localization in a narrow frequency band around $f_{sc} = 0.432$ Hz corresponds to the motion of the potential scattering source.
In contrast, a more diffuse population is distributed at $\sigma_T > 0.3$ s and spans a broad range of $f_{sc}$ values.
This indicates that these events are random noise with no specific periodic pattern.
Notably, no other significant population is distributed in the low-periodicity region ($\sigma_T < 0.3$ s), implying that scattered light noise is the dominant noise source in this data with stable and local periodicity.

%%%%%%%%%%%%%%%%%%%%%%%%
\section{Results}
This section characterizes the occurrence of scattered light noise in KAGRA using the f-cluster algorithm and quantifies its relationship with band-limited seismic motion, and derives an operational forecast threshold.
We analyzed data from April 1 to April 30, 2025.
In addition to analyzing scattered light noise using the f-cluster algorithm, we investigate the relationship between scattered light noise and the frequency band of the seismic motion.

\begin{figure}[H]
    \centering
    \includegraphics[width=1\linewidth]{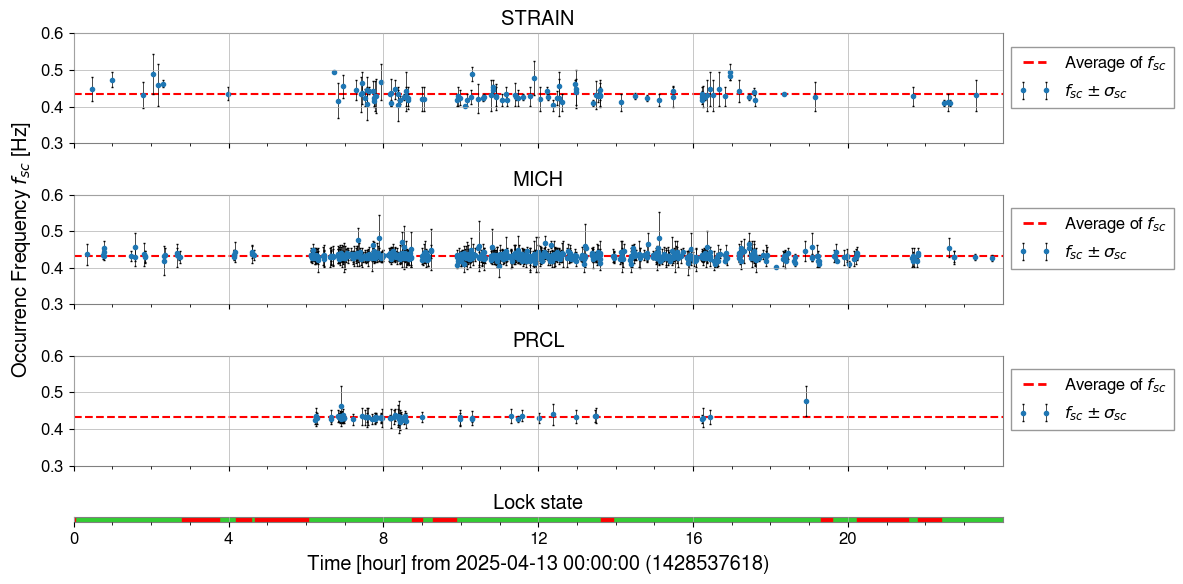}
    \caption{
    The time series of the occurrence frequencies $f_{sc}$ and standard deviations $\sigma_{sc}$ of the scattered light noise identified by the f-cluster algorithm in STRAIN, MICH, and PRCL channels, together with the interferometer lock state (green: locked, red: unlocked).
    On April 13, the average of $f_{sc}$ coincided with 0.433 Hz in each channel, as indicated by the red dashed lines.
    }
    \label{fig:1dresult}
\end{figure}
Figure \ref{fig:1dresult} shows the occurrence frequency $f_{sc}$ satisfying the condition of Eq. \ref{eq:condition} on April 13, 2025.
According to figure \ref{fig:1dresult}, the average of $f_{sc}$ in STRAIN, MICH, and PRCL is concentrated at about $0.433$ Hz, which suggests the scattered light noise in these channels shares a common source.
The MICH channel shows the strongest signatures of scattered light noise among STRAIN, MICH, and PRCL, as indicated by the largest number of SL labels assigned.
Hence, the subsequent analysis focuses on MICH to characterize the scattered light noise.

Several past studies have shown that the occurrence of scattered light noise is proportional to the magnitude of the seismic motion \cite{Accadia:2010zzb, Soni:2021, Longo:2022, Soni:2025}, and the characteristic frequency of the scattered light noise $f_\text{arch}(t)$ is modeled in Eq. \ref{eq:farch}  by the velocity of the scattering object and is validated experimentally.
Here, we consider two frequency bands of the seismic motion, $0.1 - 0.3$ Hz and $0.3 - 1.0$ Hz.
The $0.1 - 0.3$ Hz band includes the oceanic microseism peak around the 0.2 Hz \cite{Hoshino:2024}, and the $0.3 - 1.0$ Hz band contains the occurrence frequency of $f_{sc} = 0.432$ Hz.

\begin{figure}[H]
    \centering
    \begin{subfigure}[H]{\textwidth}
        \includegraphics[width=1\linewidth]{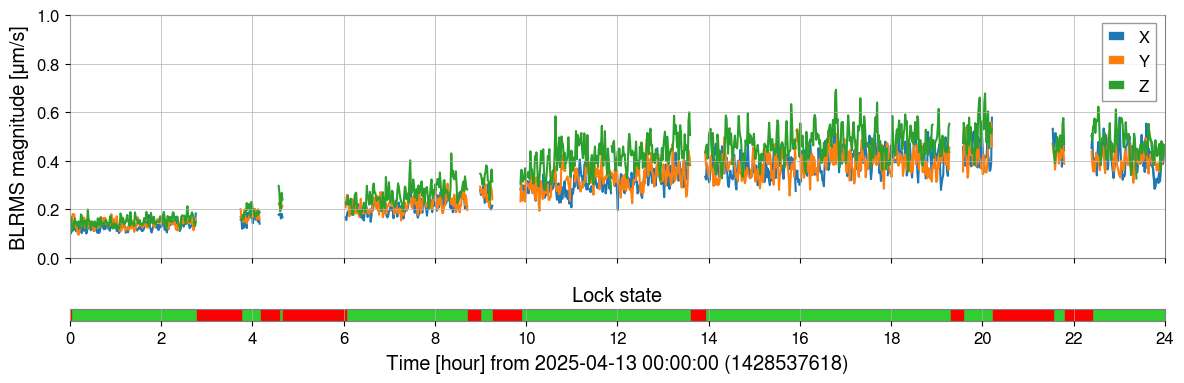}
        \caption{$0.1 - 0.3$ Hz, containing the peak of oceanic microseismic.}
        \label{fig:8a}
    \end{subfigure}
    \begin{subfigure}[H]{\textwidth}
        \includegraphics[width=1\linewidth]{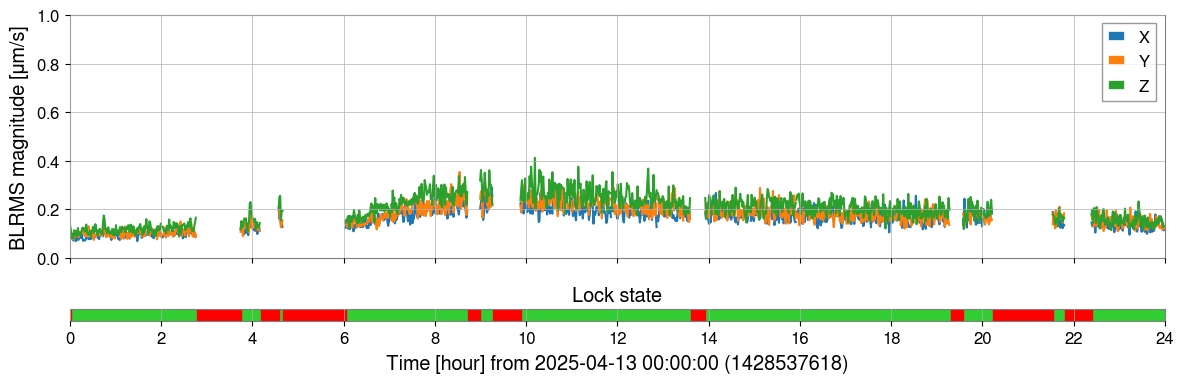}
        \caption{$0.3 - 1.0$ Hz, including the occurrence frequency of scattered light noise in KAGRA $f_{sc}=0.432$ Hz}
        \label{fig:8b}
    \end{subfigure}
    \caption{
    The band-limited RMS (BLRMS) data of the seismometer installed near the beam splitter in (a) $0.1 - 0.3$ Hz and (b) $0.3 - 1.0$ Hz bands, together with the interferometer lock state (green: locked, red: unlocked).
    The data shown is since 2025--04--13 00:00:00 UTC.
    Data are shown only when the interferometer is locked, resulting in the discontinuities.
    The typical magnitude on a day with low ground motion is around 0.1 $\mu$m/s.
    }
    \label{fig:seis}
\end{figure}

While the typical magnitude on a day with low ground motion of these bands of seismic motion is around $0.1 \mu$m/s, there is an excess of magnitude in both bands during April 13, 2025.
The magnitude in $0.1 - 0.3$ Hz band exceeded 0.2 $\mu$m/s since 02:00:00 UTC.
There is an excess of magnitude in the range of $0.3-1.0$ Hz from 08:00:00 to 20:00:00 UTC.
From figure \ref{fig:1dresult}, the occurrence of the scattered light noise is distributed non-uniformly in time.
The distribution is denser from 08:00:00 to 12:00:00.
Therefore, we proceed to investigate the relation of the scattered light noise occurrence between the seismic motion in the frequency bands of $0.1-0.3$ Hz and $0.3-1.0$ Hz.

\begin{figure}[H]
    \centering
    \begin{subfigure}[h]{0.49\textwidth}
        \includegraphics[width=1\linewidth]{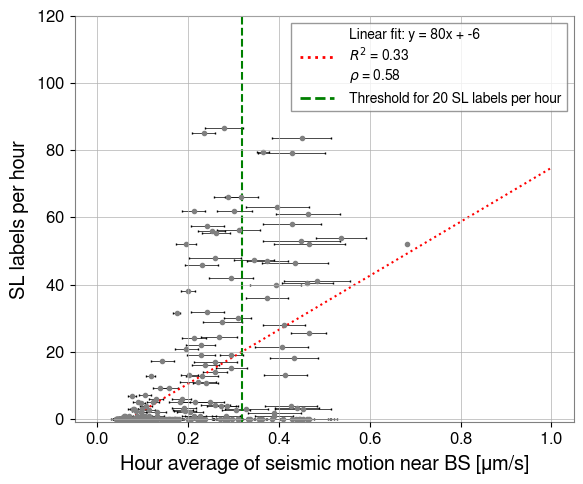}
        \caption{$0.1 - 0.3$ Hz}
        \label{fig:rate0103}
    \end{subfigure}
    \begin{subfigure}[h]{0.49\textwidth}
        \includegraphics[width=1\linewidth]{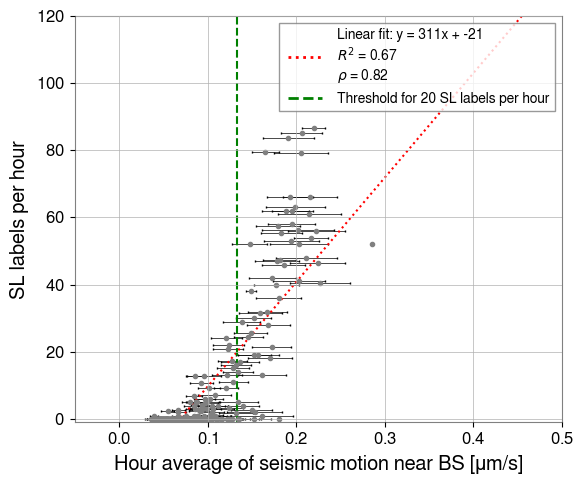}
        \caption{$0.3 - 1.0$ Hz}
        \label{fig:rate0310}
    \end{subfigure}
    \caption{
    The number of the SL labels per hour against the magnitude of the average seismic motion in the two frequency bands, $0.1 - 0.3$ Hz (left) and $0.3 - 1.0$ Hz (right).
    These data were collected from April 1, 2025, to April 30, 2025.
    The x-axis represents the one-hour average of band-limited RMS data; error bars denote the 1 $\sigma$ variability within the hour.
    The y-axis shows the hourly rate of SL labels assigned by the f-cluster algorithm.
    The red dotted line represents the fitting result obtained using the least squares method for the linear function.
    The fitting function, the coefficient of determination ($R^2$), and the Pearson correlation ($\rho$) are shown in the legend.
    The green dashed line represents the threshold of seismic motion magnitude, as determined by the hourly rate of SL labels, based on the fitting line.
    Here, we select a threshold of 20 SL labels per hour for illustration.
    }
    \label{fig:rate}
\end{figure}

From the distribution of the SL label rate shown in figure \ref{fig:rate}, only five points exceed 70 labels per hour, whereas most rates distribute around zero.
This suggests that the occurrence of scattered light noise varies over time and can be influenced by the ground motion around the interferometer.
To quantify the relationship between seismic motion and the occurrence of scattered light noise, we use the Pearson correlation coefficient.
When the Pearson correlation is close to positive or negative one, this indicates a higher linear positive or negative correlation between two sets of data.
As shown in figure \ref{fig:rate0103}, there is no apparent relation between the SL labels per hour and $0.1-0.3$ Hz seismic motion.
The Pearson correlation coefficient is 0.58, which is difficult to conclude a positive correlation.
In contrast, as shown in figure \ref{fig:rate0310}, the Pearson correlation is 0.82 for the $0.3-1.0$ Hz band of seismic motion.
This indicates a positive correlation between seismic motion and the occurrence of scattered light noise.
Therefore, we focus on the seismic motion data in the $0.3-1.0$ Hz frequency band hereafter, as the occurrence of scattered light noise in KAGRA shows a correlation with this band.

We also perform linear fitting using the least squares method on the data of scattered light occurrence against the average seismic motion, where the fitting function provides a description of the relationship.
The coefficient of determination, $R^2$, provides a measure of the performance of the fitting function by how well observed outcomes are reproduced by the fitting.
When $R^2$ is close to 1, the fitted function agrees well with the data.
If the outcome of the fitting differs from the data, $R^2$ decreases from one.
From the fitting results of figure \ref{fig:rate}, $R^2$ values are 0.33 and 0.67 for $0.1-0.3$ Hz and $0.3-1.0$ Hz seismic motion, respectively.
The linear fit between the SL label rate and seismic motion in the $0.3-1.0$ Hz band performs better than in the $0.1-0.3$ Hz band.

In the $0.3-1.0$ Hz band, when the magnitude of the average seismic motion is less than 0.2 $\mu$m/s, the data matches the linear fitting function with all the SL label rates of the data less than 40 labels per hour.
As the magnitude of the seismic motion approaches 0.2 $\mu$m/s, which is considerably larger than the typical magnitude of 0.1 $\mu$m/s, the linear fit mismatches with the data. 
The data with the SL label rate of more than 40 labels per hour are distributed around the seismic motion of 0.2 $\mu$m/s, while 0.2 $\mu$m/s is almost the largest seismic motion of this band during April 2025, except for one data point being closer to 0.3 $\mu$m/s.

From the fitting results, we can predict the occurrence rate of scattered light noise for a given level of ground motion.
By setting a target rate for the scattered light noise and solving the fitted function with this target rate, we can estimate the threshold for the ground motion.
As a demonstration, the predicted threshold for seismic motion to obtain the 20 SL labels per hour is shown by a green dashed line in figure \ref{fig:rate}.

\begin{figure}[H]
    \centering
    \includegraphics[width=1\linewidth]{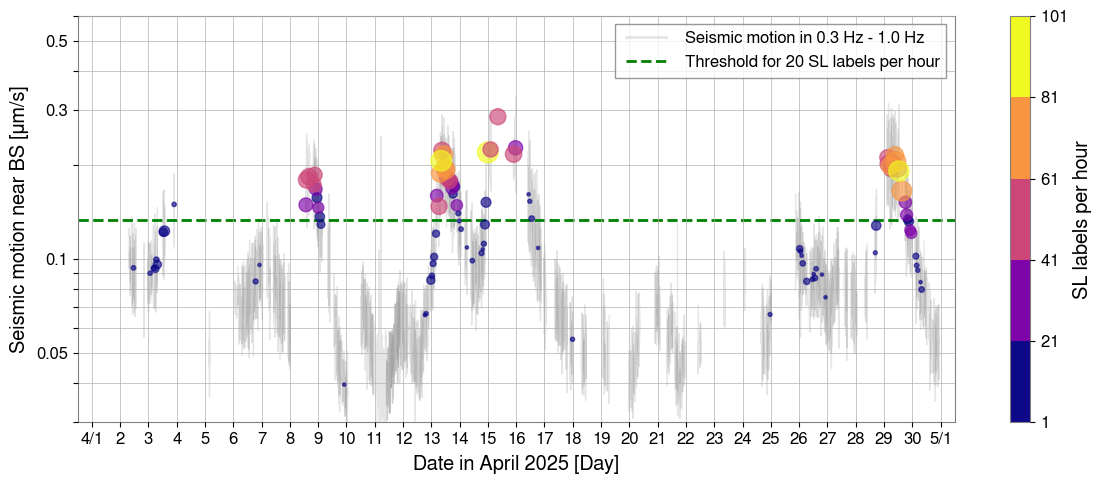}
    \caption{
    Day-to-day variation of the hourly SL labels and seismic motion amplitude for April 2025.
    The gray line represents the seismic motion between 0.3 and 1.0 Hz. 
    The colored circles display the hourly SL label rate, and their sizes are proportional to their values.
    We plotted colored dots at heights corresponding to the band-limited RMS of the seismometer data.
    Data are shown only when the interferometer is locked, resulting in discontinuities.
    The green dashed line is the 20 SL label per hour threshold from the fitting in figure \ref{fig:rate0310}.
    }
    \label{fig:daily}
\end{figure}

Figure \ref{fig:daily} is the day-to-day variation of scattered light noise occurrence in MICH and seismic motion between $0.3-1.0$ Hz in April 2025.
The number of SL labels increases when the magnitude of seismic motion in $0.3-1.0$ Hz increases.
In the past research, it was reported that the occurrence of scattered light noise is related to the magnitude of the seismic motion.
In Virgo, the scattered light noise is related to $0.1-0.3$ Hz microseismic band \cite{Longo:2023}, and it is below 0.2 Hz in LIGO \cite{Soni:2021}.
There are three peaks with the magnitude of the seismic motion greater than 0.2 $\mu$m/s on 13, 15, and 29 in April.
The data with an SL label rate of more than 70 per hour occurred on these days.
On the contrary, the seismic motion is quieter with a magnitude of around 0.05 $\mu$m/s during 10 to 12 or 17 to 25 in April. 
When the magnitude of ground motion is low, such as less than 0.05 $\mu$m/s, scattered light noise does not occur frequently.
The record from the Japan Meteorological Agency \footnote{https://www.data.jma.go.jp/kaiyou/shindan/index\_wave.html} shows that the monthly average tide height at Toyama Bay, the station closest to the KAGRA site, is 1.0 m in April 2025.
During periods of greater seismic motion, such as 13 to 16 and 29 in April 2025, the 12-hour average tide height exceeded the monthly average.
Previous research has revealed a positive correlation between seismic motion around the interferometer and ocean waves \cite{Hoshino:2024}.

From figure \ref{fig:rate0310}, we obtained the linear fitting function of the relation between the SL label rate and the seismic motion magnitude.
Hence, we also demonstrated the 20 SL label per hour threshold for forecasting the scattered light noise in the interferometer, as indicated by the green dashed line.
The threshold value of the magnitude of $0.3-1.0$ Hz seismic motion is 0.13 $\mu$m/s.
Above this threshold value, the rate of SL labels also exceeded 20 labels per hour.
This suggests a predictive method to identify periods of high scattered light contamination, even before it becomes visible.

%%%%%%%%%%%%%%%%%%%%%%%%
\section{Conclusion}
Noise in data from GW telescopes is classified into two categories: stationary noise and non-stationary noise.
Among the non-stationary noise, we focus on the scattered light noise.
Because it fluctuates across a wide frequency band within a short time, the scattered light noise can mimic the GW signal, increase the false alarm rate, or bias the parameter estimation results \cite{Powell:2018}.
There have been attempts to characterize scattered light noise by using the Hilbert-Huang transformation and empirical mode decomposition \cite{Longo:2022, Guillermo:2017}.
However, it is difficult to determine the precise frequency of the scattering object.
We proposed a new algorithm, named f-cluster, to evaluate the frequency of the scattering object and the occurrence time.
The scattered light noise shows characteristic arch-shaped features in time-frequency maps.
Therefore, the algorithm extracts the periodicity of the arch-shapes in the spectrogram without assuming the scattering object, and assigns labels that indicate scattered light noise present in the data (referred to as SL labels).

Introducing our algorithm enables us to precisely evaluate the frequency of scattered light noise.
For the demonstration, we applied our algorithm to the KAGRA data of April 2025, which contains scattered light noise of unknown origin.
The strain channel and other length degrees of freedom channels, referred to as MICH and PRCL, were analyzed. 
The occurrence frequencies of the scattered light noise were evaluated to be $f_{sc} = 0.432$ Hz, which were consistent across three channels. 
The most scattered light noise occurred in the MICH channel. 
Thus, we focused on the data of the MICH channel.

We evaluated the hour rates of the SL labels and compared them with the magnitude of seismic motion in the $0.1-0.3$ Hz and $0.3-1.0$ Hz bands.
The scatter plot of the seismometer data in the $0.3 - 1.0$ Hz band shows a linear relationship, with a Pearson correlation coefficient of 0.82 within this band. 
This suggests that the seismic motion in $0.3 - 1.0$ Hz band is associated with the scattering source of $f_{sc} = 0.432$ Hz.

From the fitting result of the scatterplot for $0.3 - 1.0$ band, we can predict the occurrence rate of scattered light noise for a given level of ground motion. 
As a demonstration, we selected the threshold for seismic motion to obtain the 20 SL labels per hour. 
We confirmed that the evaluated threshold is possible to forecast the occurrence of the scattered light noise, as seen in the day-to-day variation of April 2025. 

From these results, our algorithm is capable of evaluating the frequency of scattered light noise and the occurrence time, without assuming the scattering object.
Therefore, our algorithm can serve as a new tool for characterizing scattered light noise.
It is expected that we can apply the algorithm to LIGO and Virgo data for the identification of the scattered light noise.

The occurrence time and frequency of scattering light noise, which the f-cluster algorithm can provide, are helpful in identifying the physical sources of the scattering object in noise hunting.
Correlation analyses using auxiliary channels \cite{Essick:2020cyv, Essick:2021,iDQ:2024, Smith:2011an, PhysRevD.94.042004} can support the noise hunting to specify the source of the scattered light noise.

\section*{Acknowledgement}

This work was supported in part by the National Science and Technology Council, Taiwan, under Grant No. NSTC 114-2112-M-A49-001.

This work was supported in part by the joint research program of the Institute for Cosmic Ray Research (ICRR), The University of Tokyo. This work was supported by JSPS KAKENHI Grant Number JP23K13120.

The authors Chia-Jui Chou and Yi Yang are supported by the start-up grant (2025F0201-000-02 and 2025F0201-000-04) from ShanghaiTech University.

\section*{References}
\bibliographystyle{unsrt}
\bibliography{SL_reference}

\end{document}